\documentclass[longauth]{aa}  
\usepackage{graphicx}
\usepackage{txfonts}

\newcommand{\orcid}[1]{\href{https://orcid.org/#1}{\includegraphics[width=10pt]{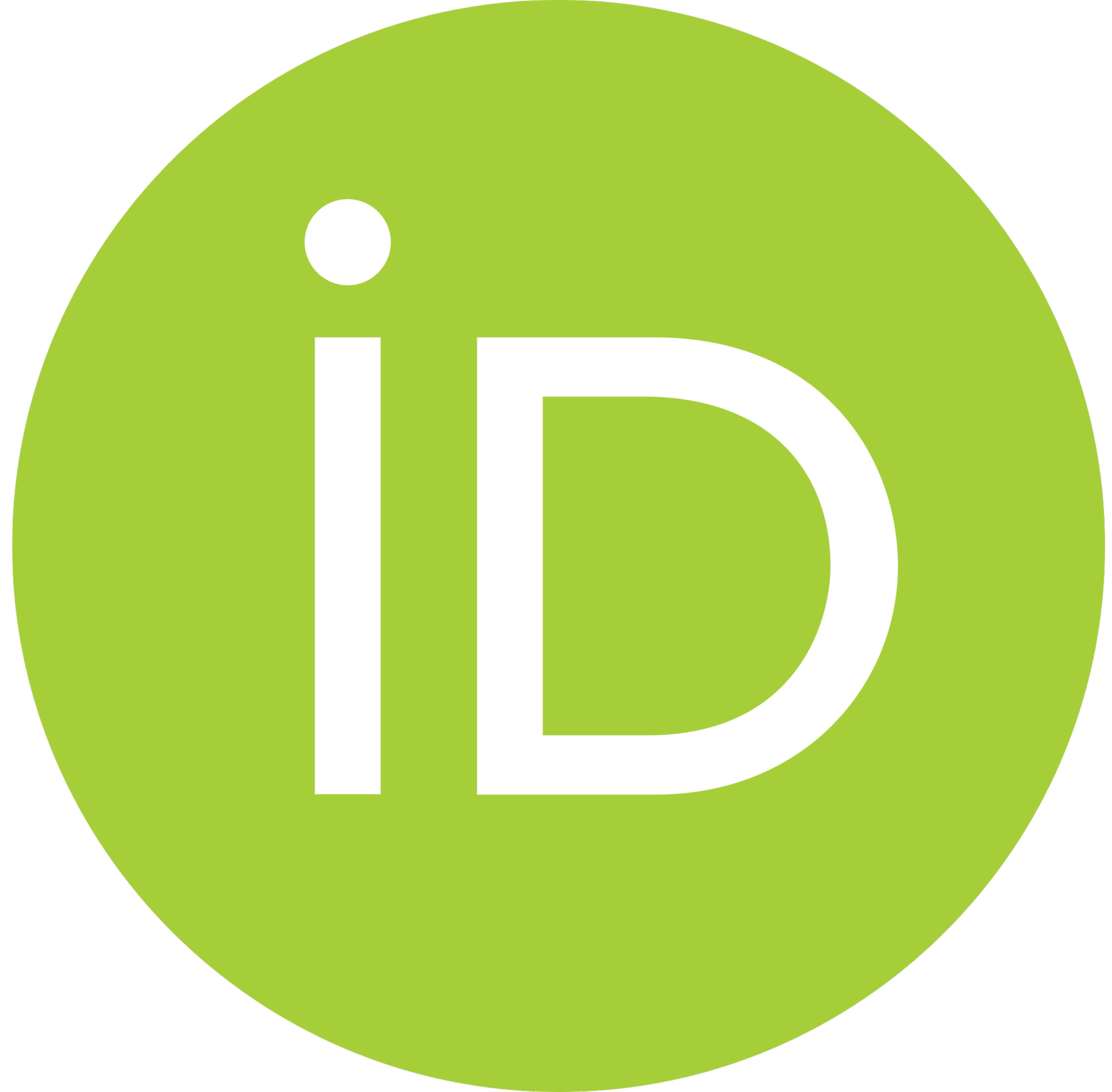}}}

\usepackage{hyperref}      
\hypersetup{colorlinks=true,linkcolor=blue,citecolor=blue,filecolor=blue,urlcolor=blue}

\newcommand{\HII}{\ion{H}{II}}

\newcommand{\kms}{\rm km\,s$^{-1}$}
\def \kms {{\rm km\,s$^{-1}$}}

\def \msun {{$\mathrm{M}_\odot$}}

\def \micron{\hbox{$\mu$m}}

\def \cmcb {{\rm cm$^{-3}$}}

\def \ntdp {{\rm N$_2$D$^+$}}

\newcommand{\LEt}[1]{}

\newcommand{\aifa}{Argelander-Institut f\"{u}r Astronomie, Universit\"{a}t Bonn, Auf dem H\"{u}gel 71, 53121, Bonn, Germany}
\newcommand{\eso}{European Southern Observatory (ESO), Karl-Schwarzschild-Stra{\ss}e 2, 85748 Garching, Germany}
\newcommand{\eao}{East Asian Observatory, 660 N. A‘ohok¯u Place, University Park, Hilo, HI, USA}
\newcommand{\cfa}{Center for Astrophysics | Harvard \& Smithsonian, Cambridge, MA, USA}
\newcommand{\chalmers}{Dept. of Space, Earth and Environment, Chalmers University of Technology, SE-412 96 Gothenburg, Sweden}
\newcommand{\univ}{Department of Astronomy, University of Virginia, 530 McCormick Road Charlottesville, 22904-4325 USA}
\newcommand{\kalvi}{Kavli Institute for Astronomy and Astrophysics, Peking University, 5 Yiheyuan Road, Haidian District, Beijing 100871, China}
\newcommand{\mpe}{Max-Planck-Institut f\"{u}r extraterrestrische Physik, Giessenbachstrasse 1, 85748 Garching bei M\"{u}nchen, Germany}
\newcommand{\ipac}{IPAC, Mail Code 100-22, Caltech, 1200 E. California Boulevard, Pasadena, CA 91125, USA}
\newcommand{\ushef}{Department of Physics and Astronomy, The University of Sheffield, Hicks Building, Hounsfield Road, Sheffield, S3 7RH, UK}
\newcommand{\mpia}{Max Planck Institute for Astronomy, K\"{o}nigstuhl 17, D-69117 Heidelberg, Germany}
\newcommand{\ljmu}{Astrophysics Research Institute, Liverpool John Moores University, 146 Brownlow Hill, Liverpool L3 5RF, UK}
\newcommand{\csicinta}{Centro de Astrobiolog\'{i}a (CSIC/INTA), Instituto Nacional de T\'{e}cnica Aeroespacial, 28850 Torrej\'{o}n de Ardoz, Madrid, Spain}
\newcommand{\ign}{Observatorio Astron\'omico Nacional (OAN, IGN), Calle Alfonso XII 3, 28014, Madrid, Spain}
\newcommand{\inaf}{INAF Osservatorio Astrofisico di Arcetri, Largo E. Fermi 5, 50125 Florence, Italy}

\begin{document} 

\title{Mother of dragons}

\subtitle{A massive, quiescent core in the dragon cloud (IRDC G028.37+00.07)}

\authorrunning{A.~T.~Barnes et al. }

\author{A.~T.~Barnes\,\orcid{0000-0003-0410-4504},\inst{1,2}\thanks{\url{ashleybarnes.astro@gmail.com}} 
        J.~Liu,\inst{3}
        Q.~Zhang,\inst{4}
        J.~C.~Tan,\inst{5,6}
        F.~Bigiel,\inst{2}
        P.~Caselli,\inst{7}
        G.~Cosentino,\inst{5}
        F.~Fontani,\inst{8}
        J.~D.~Henshaw\,\orcid{0000-0001-9656-7682},\inst{9,10}
        I.~Jiménez-Serra,\inst{11}
        D-S.~Kalb,\inst{10}
        C.~Y.~Law,\inst{1,5}
        S.~N.~Longmore,\inst{9} 
        R.~J.~Parker,\inst{12}\thanks{Royal Society Dorothy Hodgkin Fellow}
        J.~E.~Pineda,\inst{7}
        A.~S\'anchez-Monge\,\orcid{0000-0002-3078-9482},\inst{13}
        W.~Lim,\inst{14}
        K.~Wang\inst{15}
      }

\institute{\eso \and
            \aifa \and 
            \eao \and
            \cfa \and 
            \chalmers \and 
            \univ \and
            \mpe \and
            \inaf \and
            \ljmu \and
            \mpia \and
            \csicinta \and
            \ushef \and
            \ign \and
            \ipac \and
            \kalvi
            }

\date{Received 12/12/2022; accepted 08/03/2023}

 
\abstract{
\LEt{ General notes: A.) I have edited to UK English spelling and grammar conventions. B.) A\&A uses the past tense to describe specific methods used in a paper and the present tense to describe general methods as well as findings, including the findings of recent papers (within the past ten or so years). Kindly make any necessary changes (I have made some, but my edits are by no means exhaustive in this respect). See Sect. 6 of the language guide https://www.aanda.org/for-authors/language-editing/6-verb-tenses.}Core accretion models of massive star formation require the existence of massive, starless cores within molecular clouds. 
Yet, only a small number of candidates for such truly massive, monolithic cores are currently known.}
{Here we analyse a massive core in the well-studied infrared-dark cloud (IRDC) called the `dragon cloud' (also known as G028.37+00.07 or `Cloud C'). 
This core (C2c1) sits at the end of a chain of a roughly equally spaced actively star-forming cores near the center of the IRDC.}
%
{We present new high-angular-resolution 1\,mm ALMA dust continuum and molecular line observations of the massive core.}
{The high-angular-resolution observations show that this region fragments into two cores, C2c1a and C2c1b, which retain significant background-subtracted masses of 23\,\msun\ and 2\,\msun\ (31\,\msun\ and 6\,\msun\ without background subtraction), respectively. The cores do not appear to fragment further on the scales of our highest-angular-resolution images (0.2\arcsec, 0.005\,pc\,$\sim$\,1000\,AU). We find that these cores are very dense ($n_\mathrm{H_2}>10^6$\,cm$^{-3}$) and have only trans-sonic non-thermal motions ($\mathcal M_\mathrm{s}\sim1$). 
Together the mass, density, and internal motions imply a virial parameter of $<$\,1, which suggests the cores are gravitationally unstable, 
unless supported by strong magnetic fields with strengths of $\sim$1-10\,mG. From CO line observations, we find that there is tentative evidence for a weak molecular outflow towards the lower-mass core, and yet the more massive core remains devoid of any star formation indicators.}
{We present evidence for the existence of a massive, pre-stellar core, which has implications for theories of massive star formation. 
This source warrants follow-up higher-angular-resolution observations to further assess its monolithic and pre-stellar nature.
}
  

   \maketitle
%
\section{Introduction}
\label{sec_int}

\begin{figure*}
    \centering
        \includegraphics[width=\textwidth]{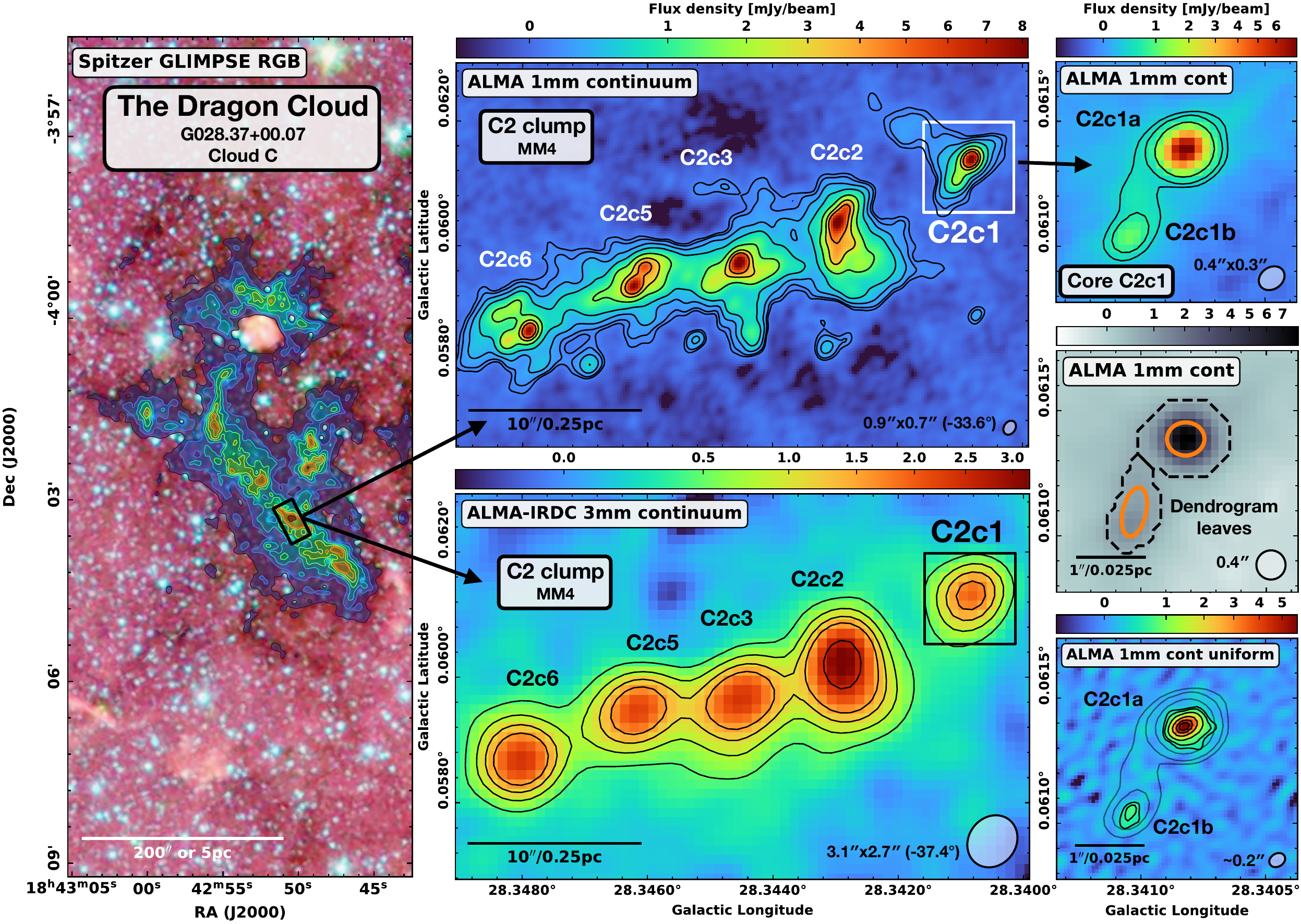}
        \vspace{-5mm}
    \caption{{Overview of the dragon cloud (also known as G028.37+00.07 or Cloud C), the C2 clump (also known as MM2), and the C2c1 core region.} \LEt{ Verify that your intended meaning has not been changed.}({\it left}) Three-colour image of the Galactic plane in which the IRDC Cloud C can be seen as a strong dark extinction feature. In this image, red is 8\,\micron, green is 5.8\,\micron,\ and blue is 4.5\,\micron\ emission from the {\it Spitzer} GLIMPSE \LEt{ Consider defining.}survey \citep{churchwell_2009}. Overlaid as coloured contours is the combined near- and mid-infrared extinction-derived mass surface density map, in levels of 0.1, 0.15, 0.2, 0.25, 0.3, 0.35, 0.375, and 0.45\,g\,cm$^{-2}$ \citep{Kainulainen2013}. ({\it center top}) ALMA 1\,mm dust continuum at $0.8$\arcsec\ resolution \citep{LiuJ2020}, overlaid with contours of 0.084, 0.14, 0.28, 0.56, 1.4, 2.8, and 4.2\,mJy/beam. ({\it centre bottom}) ALMA 3\,mm dust continuum taken from ALMA-IRDC at $3$\arcsec\ resolution \citep{Barnes2021_almairdc}, overlaid with contours of 0.49, 0.81, 1.13, 1.46, and 2.43\,mJy/beam. ({\it right top}) Zoom in on the high-resolution 1\,mm continuum image taken with only the longest baseline configuration, overlaid with contours of 0.4, 0.6, and 1.2\,mJy/beam. ({\it right center}) ALMA 1\,mm dust continuum with a circularised $0.4$\arcsec\ beam, which was used for the dendrogram analysis. The dashed black contours indicate the boundaries of the identified cores, whilst the orange ellipses show the intensity-weighted second moment along the two spatial dimensions within the contour (e.g. \citealp{rosolowsky_2008}). ({\it right bottom}) Same 1\,mm observations imaged with a robust -2 (close to uniform) weighting scheme to maximise the resolution (0.24\arcsec\ x 0.19\arcsec, or $\sim$0.05\,pc), overlaid with contours of 0.4, 0.6, 1.2, 2.5, 4, and 5\,mJy/beam (solid black line). Shown on the right of the ALMA observation panels is the beam size, and a scale bar is shown in the lower left of all panels.}
    \label{fig1}
\end{figure*}

Massive stars (i.e. with initial masses $>8$\,\msun) inject vast amounts of energy and momentum into the interstellar medium during and at the end of their relatively short lives, thus helping drive both local and galaxy-scale physical and chemical evolution \citep[e.g.][]{krumholz_2014}. 
However, understanding how massive stars form remains one of the major unanswered questions in astrophysics \citep[e.g.][]{Tan2014}. 

One way to test different formation theories is to study the initial conditions of massive star formation. 
In particular, `core accretion' theories (e.g. \citealp{McLaughlin1997,mckee_2003}) invoke the existence of massive, gravitationally bound pre-stellar cores as the initial condition. On the other hand, `competitive accretion' theories and simulations (e.g. \citealp{bonnell_2001, Wang2010,Padoan2020}) do not involve such structures, but rather start with low-mass protostellar cores, some of which competitively accrete to form high-mass stars, being fed by global infall from the protocluster clump.
Other authors have pointed out that the gas around massive protostars and star clusters often exhibits a hub-filament type morphology (e.g. \citealp{Myers2009}); this has implications for how material is spatially and kinematically distributed around protostellar systems but in itself does not distinguish between the two basic scenarios.

Thus, identifying and characterising massive pre-stellar cores remains a primary way to distinguish between massive star formation scenarios.
However, to date, only a handful of candidate massive pre-stellar cores have been identified. A non-exhaustive list of candidates and/or searches includes: \citet{Bontemps2010}, \citet{Tan2013}, \citet{Duarte-Cabral2013}, \citet{Wang2014}, \citet{Cyganowski2014}, \citet{Sanhueza2017}, \citet{Nony2018}, \citet{Louvet2019}, and \citet{Kong2018}. 
Yet, despite these efforts, there are relatively few candidates of massive pre-stellar cores. 
It thus remains important to try to identify additional examples of massive pre-stellar cores

Recently, a study of high-sensitivity 3\,mm dust continuum and molecular line ALMA \LEt{ Consider defining.}observations at $\sim$\,2\arcsec\ resolution identified a population of compact, massive cores that presented the potential birth sites of high-mass star formation (\citealp{Barnes2021_almairdc, Fontani2021}). 
Out of this sample of 19 cores, one stood out as being a particularly strong monolithic core candidate.\  This core, known as C2c1 (called core\,5 in \citealp{LiuJ2020}), is located within the `dragon cloud' (also known as G028.37+00.07 and referred to as Cloud C by \citealt{butler_2009}; see \citealp{Wang2015}) at a distance of 5\,kpc (\citealp{Simon2006b}). 
\citet{Barnes2021_almairdc} highlighted that this core does not fragment down to $\sim$\,0.01\,pc, and at this scale it still has more than enough mass to form a high-mass star (total mass of 40\,M$_\odot$; also see \citealp{Wang2011, Zhang2015}). 
Moreover, this source shows no signs of infrared point source emission  (see our Fig.\,\ref{fig1} and Sect. \,\ref{sec_sf}; \citealt{Kong2019, Barnes2021_almairdc}). 
Broadband spectral line imaging shows that C2c1 is associated with very little molecular line emission \citep{Zhang2015}.
It has only low levels of CO emission, indicating a CO depletion factor (fraction of CO molecules frozen out of the gas phase onto dust grains) of up to 10$^3$ at a scale of over 0.07\,pc. 
In addition, the region also shows moderate N$_2$D$^+$ emission, highlighting an enhanced abundance of deuterated molecules, which is typically only seen towards the coldest and densest environments (with elevated CO freeze-out; see e.g. \citealp{kong_2016, barnes_2016}).   
Altogether, this is indicative of a pre-stellar nature for core C2c1 \citep{Wang2011, Zhang2015, kong_2016, Kong2017, LiuJ2020} and makes it one of only a handful of potential monolithic massive core candidates. 

Due to its location at the end of a string of cores near the centre of Cloud C, and its positioning at the edge of the primary beam response for most of the initial interferometric observations (e.g. \citealp{Wang2011, Zhang2015, kong_2016, Kong2017}),  C2c1 has mostly eluded detailed study (e.g. \citealp{LiuJ2020, Barnes2021_almairdc}). 
In this work, we make use of new high-resolution continuum and \ntdp\  observations (the latter a sensitive molecular line tracer of dense and cold gas) to study C2c1 in detail, in an effort to unveil its geometrical, chemical, and dynamical structure. 
The properties of C2c1 determined in this work are summarised in Table\,\ref{tab1}. 

\begin{table}
\caption{{Summary of the C2c1a and C2c1b core properties.} We outline how each of these properties is measured in Sects. \ref{sec_frag}, \ref{sec_dyn}, and \ref{sec_stab}. The quoted $T_\mathrm{dust}$ values are taken from \citet{Wang2012}. Note that all the properties listed here that are calculated with the mass use the values that include background subtraction.}              
\label{tab1}      
\centering                                      
\begin{tabular}{l c c}          
\hline\hline                        
Property &  C2c1a & C2c1b \\
\hline                  

Longitude [deg]  & 28.3408 & 28.3410 \\
Latitude [deg] & 0.0614 & 0.0611 \\
$r_\mathrm{eff}$ [arcsec] & 0.57 & 0.43 \\
$r_\mathrm{eff}$ [pc] & 0.014 & 0.010 \\
$r_\mathrm{eff}$ [AU] & 2827 & 2127 \\
$S_v$ (no background sub.) [mJy] & 12.89$\pm$1.29 & 2.43$\pm$0.24 \\
$S_v$ (background sub.) [mJy] & 9.65$\pm$0.97 & 0.64$\pm$0.06 \\
$T_\mathrm{dust}$ [K] & 10.4$\pm$3.0 & 10.4$\pm$3.0 \\
$M$ (no background sub.) [M$_\odot$] & 30.9$\pm$12.4 & 5.8$\pm$2.3 \\
$M$ (background sub.) [M$_\odot$] & 23.1$\pm$9.2 & 1.5$\pm$0.6 \\
$n_\mathrm{H_2}$ [10$^{6}$\,cm$^{-3}$] & 31.21$\pm$12.48 & 4.86$\pm$1.94 \\
$t_\mathrm{ff}$ [yr] & 5525$\pm$2762 & 14000$\pm$7000 \\
$v_\mathrm{LSR}$ [km\,s$^{-1}$] & 79.36$\pm$0.03 & 79.08$\pm$0.04 \\
$\sigma_\mathrm{obs}$ [km\,s$^{-1}$] & 0.35$\pm$0.04 & 0.25$\pm$0.05 \\
$c_\mathrm{s}$ [km\,s$^{-1}$] & 0.19$\pm$0.06 & 0.19$\pm$0.06 \\
$\sigma_\mathrm{obs,corr}$ [km\,s$^{-1}$] & 0.34$\pm$0.04 & 0.25$\pm$0.05 \\
$\sigma_\mathrm{NT}$ [km\,s$^{-1}$] & 0.34$\pm$0.10 & 0.24$\pm$0.07 \\
$\sigma_\mathrm{tot}$ [km\,s$^{-1}$] & 0.39$\pm$0.12 & 0.31$\pm$0.09 \\
$\mathcal M_\mathrm{s}\,=\,\sigma _\mathrm{NT} / c_\mathrm{s}$ & 1.8$\pm$0.7 & 1.2$\pm$0.5 \\
$M_\mathrm{J}$ [M$_\odot$] & 0.05$\pm$0.03 & 0.13$\pm$0.06 \\
$\lambda_\mathrm{J}$ [AU] & 733$\pm$366 & 1858$\pm$929 \\
$\alpha_{\rm vir}$ & 0.10$\pm$0.05 & 0.74$\pm$0.37 \\
$M_\mathrm{J,tot}$ [M$_\odot$] & 0.42$\pm$0.21 & 0.52$\pm$0.26 \\
$B$ [mG] & 10.0$\pm$5.0 & 1.0$\pm$0.5 \\
$B_\mathrm{med}$ [mG] & 14.4$\pm$7.2 & 4.3$\pm$2.2 \\
$\Sigma_\mathrm{cl}$ [g\,cm$^{-2}$] & 0.62$\pm$0.25 & 0.62$\pm$0.25 \\
$\sigma_\mathrm{c,vir}$ [km\,s$^{-1}$] & 0.70$\pm$0.35 & 0.35$\pm$0.18 \\
$\sigma_\mathrm{c,vir}(B_\mathrm{med})$ [km\,s$^{-1}$] & 0.26$\pm$0.13 & 0.14$\pm$0.07 \\
$R_\sigma = \sigma_\mathrm{tot}/\sigma_\mathrm{c,vir}$ & 0.56$\pm$0.28 & 0.87$\pm$0.43 \\
$R_\mathrm{\sigma(Bmed)} = \sigma_\mathrm{tot}/\sigma_\mathrm{c,vir}(B_\mathrm{med})$ & 1.50$\pm$0.75 & 2.26$\pm$1.13 \\

\hline                                         
\end{tabular}
\end{table}

%
\section{Observations}
\label{sec_obs}

To investigate the dense gas properties within Cloud C, we collected several high-angular-resolution dust continuum and molecular line observations with ALMA. 
The data were taken as part of the project 2017.1.00793.S (see \citealp{LiuJ2020} for the complete data reduction details and Liu et al. in prep). 
In our analysis, we made use of the 1\,mm continuum images with the observed longest baseline configuration, which provides a beam of 0.396\arcsec\,$\times$\,0.314\arcsec\ with a position angle of 55$^{\circ}$, and a noise of 0.1\,mJy\,beam$^{-1}$. 
The CO(2-1) cube (230.538\,GHz rest frequency) also imaged with this configuration has a beam of 0.40\arcsec\,$\times$\,0.32\arcsec\ with a position angle of 60.8$^{\circ}$, a spectral resolution of 0.16\,\kms, and noise of $\sim$4\,mJy\,beam$^{-1}$ per channel.
For the \ntdp\,(3-2) cube (231.322\,GHz rest frequency), we made use of the images that include both array configurations observed as part of project 2017.1.00793.S, as the smaller baselines were then needed to recover the lower brightness emission.
They have a beam of 0.93\arcsec\,$\times$\,0.74\arcsec\ with a position angle of -3.4$^{\circ}$, a spectral resolution of 0.16\,\kms, and noise of 0.3\,K per channel.

\section{Fragmentation and mass distribution of C2c1}
\label{sec_frag}

Figure\,\ref{fig1} shows how the dust extinction and continuum emission (i.e. mass) distribution of the region changes from scales of several parsecs (cloud)  down to 0.01 parsec (core). 
We see that the initially filamentary cloud (left panels) breaks down into regularly spaced cores within the MM4 `clump' region (centre panels). 
The brightest cores within this region have been studied in detail (e.g. \citealp{Zhang2009, Zhang2015, Wang2008, Wang2011, Wang2012}); they are found to be relatively massive but to already show signs of near-infrared emission and SiO and CO outflows -- indicative of active star formation. 
The core C2c1 is seen at the end of this string of cores and appears to be somewhat disconnected in the extended envelope of millimetre-continuum emission that surrounds the main cores. 
A more detailed look at C2c1 (right panels) shows that it in fact breaks into two cores surrounded by a common envelope -- denoted here as C2c1a and C2c1b (see also \citealp{Zhang2015}).  

We characterised the mass fragmentation of this region using a dendrogram analysis \citep{rosolowsky_2008}, which was chosen to allow a direct comparison to other works that study our cloud sample \citep{Henshaw2016_frag,Henshaw2017_h6alma,liu_2018, Barnes2021_almairdc}. 
We ran the dendrogram analysis using the continuum maps with a circularised $\sim$\,0.4\arcsec\ beam, which was corrected for the primary beam response. 
We note, however, that the primary beam correction has little effect on the total flux or flux distribution towards and around the core, as the full mosaic covers a much larger region (i.e. the right panels of Fig.\,\ref{fig1} only show a zoomed-in view towards the core).
The set of parameters that are used for the determination of the dendrogram structure is: {\sc min\_value} = 3\,$\sigma$=0.22\,mJy\,beam$^{-1}$ (the minimum intensity considered in the analysis); {\sc min\_delta} = 3\,$\sigma$ (the minimum spacing between isocontours); and {\sc min\_pix} = 1\,beam area $\sim$ 12 pixels (the minimum number of pixels contained within a structure). 
We found that the identified structure was robust against changes ($\pm2\sigma$) in the dendrogram parameters; core C2c1a has a peak of $\sim60\sigma$.

We find that the dendrogram analysis cleanly identifies both cores within the C2c1 region as `leaves' --  the highest level (smallest) structures in the dendrogram hierarchy. \LEt{ We do not allow the use of "e.g." or "i.e." within the main text (in parentheses or within figure/table captions is fine).}
These cores are overlaid as dashed black contours on the continuum map shown in Fig.\,\ref{fig1} (right centre panel).
We find the effective radii ($r_\mathrm{eff} = \sqrt{A/\pi}$, where $A$ is the area enclosed within the dendrogram boundary) of C2c1a and C2c1b to be 0.57\arcsec and 0.43\arcsec, which, at the cloud kinematic distance ($d$) of 5\,kpc \citep{Simon2006b}, corresponds to 0.014\,pc (2800\,AU) and 0.010\,pc (2100\,AU), respectively.

The flux densities ($S_\nu$) integrated within C2c1a and C2c1b are 12.9 and 2.4\,mJy.\ After subtracting the background level, following the method outlined in \citet{Barnes2021_almairdc}, they are 9.7 and 0.6\,mJy (75 and 25\,$\%$ of the flux before background subtraction). We made use of these flux values in the following mass estimation,
\begin{equation}
    M = \frac{d^{2} S_\nu R_\mathrm{gd}}{\kappa_\nu B_\nu(T_\mathrm{dust})}, 
\label{equ:mass}
\end{equation}
where $R_\mathrm{gd} = 141$ is the ratio of the total dust mass (i.e. the gas plus the dust) over the refractory-component dust mass \LEt{ Verify that your intended meaning has not been changed.}(assuming a typical interstellar composition of H, He, and metals; \citealp{draine11}),
$B_\nu(T)$ is the Planck function for a given temperature ($T_\mathrm{dust}$) at a representative frequency of $\nu$\,=\,216.44\,GHz, and $\kappa_\nu = \,\kappa_0 \left( \nu / \nu_{0} \right)^\beta \approx 0.175$\,cm$^2$g$^{-1}$, when assuming the $\nu_{0}$\,=\,230\,GHz, $\beta = 1.75,$ and $\kappa_0$\,=\,0.899\,cm$^{2}$\,g$^{-1}$ obtained from \citealp{ossenkopf_1994} for an \citet[i.e. MNR,][]{mathis_1977} size distribution with thin ice mantles after $10^{5}$\,yr of coagulation at a density of 10$^{6}$\,cm$^{-3}$. 
\citet{Wang2012} used $2.8^{\prime\prime}$ NH$_3$ (1,1) and (2,2) observations from the Very Large Array (VLA) to obtain rotational temperatures of 9.2\,K and 11.6\,K towards C2c1a and C2c1b (also see Table\,1 of \citealp{Zhang2015}).
As the cores are not fully resolved in these VLA observations, we took the mean temperature of 10.4\,K for $T_\mathrm{dust}$ in this analysis.
When doing so, we calculated background-subtracted masses for C2c1a and C2c1b of 23.1\,\msun\ and 1.5\,\msun\ (31\,\msun\ and 6\,\msun\ without the background subtraction).


The molecular hydrogen number density of each core was determined assuming a uniform density sphere, 
\begin{equation}
    n_\mathrm{H_2} = \frac{M}{\frac{4}{3} \pi r_\mathrm{eff}^3 \mu_\mathrm{H_2} m_\mathrm{H}},
\end{equation}
where $\mu_\mathrm{H_2} = 2.8$ is the mean molecular weight per hydrogen molecule, and $m_\mathrm{H}$ is the mass of a hydrogen atom.
We find $n_\mathrm{H_2}$ 31.2\,$\times10^6$\,cm$^{-3}$ and 4.9\,$\times10^6$\,cm$^{-3}$ for C2c1a and C2c1b. 
The corresponding local free-fall time is calculated as
\begin{equation}
    t_\mathrm{ff} = \left( \frac{\pi^2 r_\mathrm{eff}^3}{8 GM } \right ) ^{0.5} = \left( \frac{3 \pi}{32 G \mu_\mathrm{H_2} m_\mathrm{H} n_\mathrm{H_2}} \right ) ^{0.5},
\end{equation}
where $G$ is the gravitational constant. 
We find local free-fall times of 5500 and 14000\,years for C2c1a and C2c1b, respectively.

Overall, these properties indicate that both cores, C2c1a and C2c1b, are dense, have very short free-fall times, and remain un-fragmented on scales of a few thousand AU. 
In particular, they highlight C2c1a as a strong candidate to form a massive star ($>$8\,\msun). 
That said, there are several uncertainties in our calculations that are worth outlining. 
We assumed a typically $\sim$\,10 per cent uncertainty in the absolute flux scale of the ALMA observations, and, following \citet{Sanhueza2017}, we assumed an uncertainty of $\sim$\,30 per cent dust opacity. 
These uncertainties in the dust opacity, dust emission fluxes, and a $\sim$\,30 per cent uncertainty on the distance propagate to give an uncertainty of 40 to 50 per cent for the masses.
Moreover, we note that choosing the average {\it Herschel}-derived dust temperature across the region of $\sim$\,15\,K (determined at much lower angular resolution; \citealp{Barnes2021_almairdc}), and a lower $R_\mathrm{gd} = 100$, gives background-subtracted masses of 9.5\,\msun\ and 0.6\,\msun\ for C2c1a and C2c1b.
While only representing lower limits for each core, these masses further highlight that even in this lower limit case the core C2c1a contains enough mass to form a high-mass star ($>$8\,\msun). 

%
\section{Gas dynamics of C2c1}
\label{sec_dyn}

\begin{figure}
    \centering
        \includegraphics[width=\columnwidth]{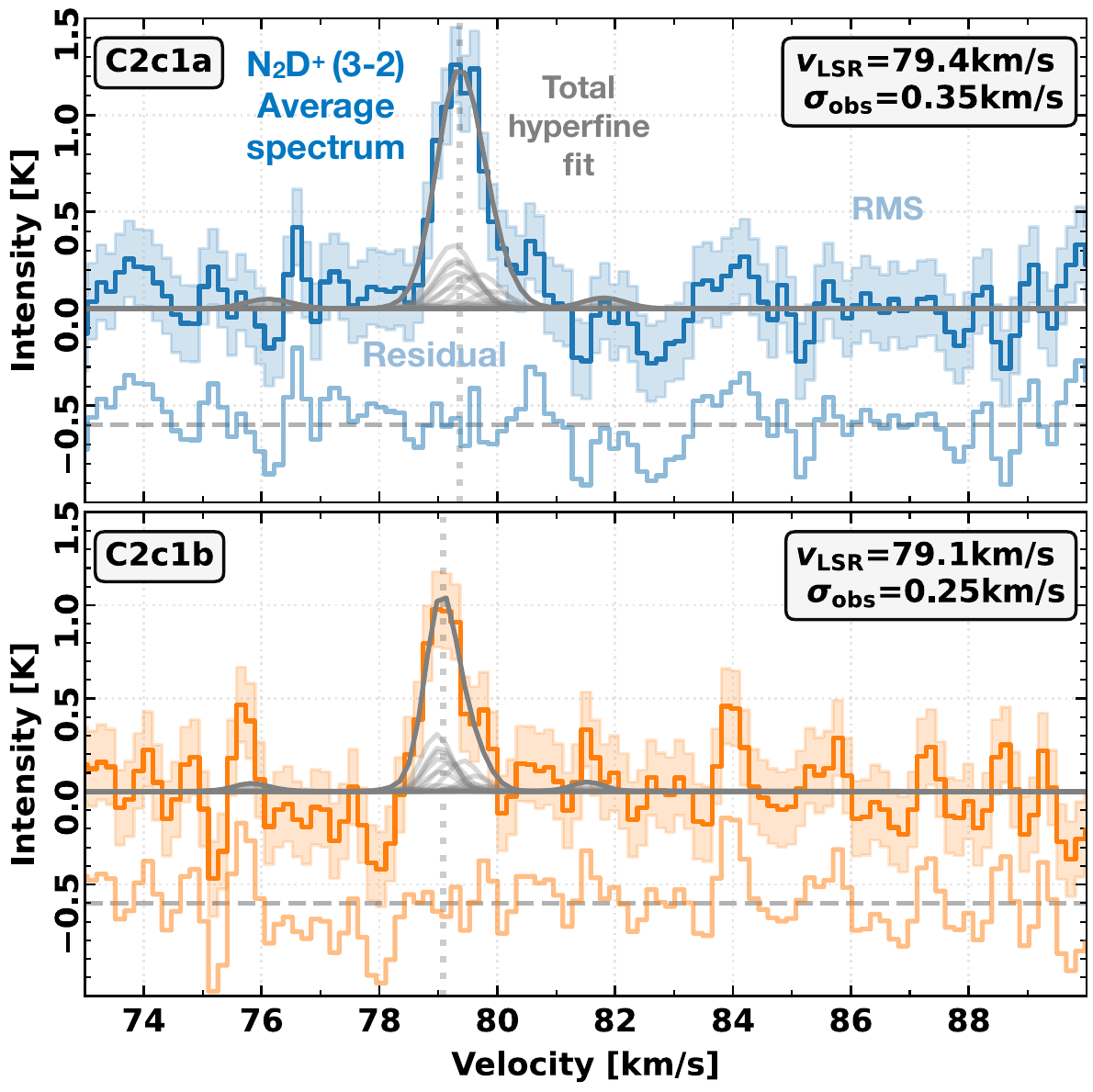}
    \caption{{Spectrum of N$_2$D$^+$\,(3--2) towards C2c1a (upper panel) and C2c1b (bottom panel).} Shown as coloured lines is the spectrum averaged within the boundary of the cores (see Fig.\,\ref{fig1}), and the coloured shaded regions indicate the uncertainty (rms). Shown in grey is the \ntdp\,(3-2) hyperfine fit to the spectrum assuming optically thin ($\tau=0.1$) conditions (faded grey curves show the individual hyperfine components of the fit), with the fit parameters shown in the upper right of each panel. The faded spectrum below each profile is the residual of the fit (offset by $-0.6$\,K). Figures\,\ref{fig_nthp_moms} and \ref{fig_nthp_channelmaps} show moment and channel maps of the \ntdp\,(3-2) emission across the region, respectively.}
    \label{fig_spec}
\end{figure}

We determined the dynamical properties of both cores using this \ntdp\,(3-2) emission (see Appendix \ref{sec_app_n2d}).
Figure\,\ref{fig_spec} shows the mean spectra for both cores, which have been extracted within the boundary defined by the dendrogram leaf contours.
We fit the spectra with the full hyperfine structure of the \ntdp\,(3-2) line assuming optically thin ($\tau=0.1$) conditions using the {\sc PySpecKit} package \citep{Ginsburg2011,Ginsburg2022}. 
We corrected the measured velocity dispersion ($\sigma_\mathrm{obs}$) for the contribution of the velocity resolution ($\sigma_\mathrm{res}$),   
\begin{equation}
    \sigma_\mathrm{obs,corr}^2 = \sigma_\mathrm{obs} ^2 - \sigma_\mathrm{res} ^2 = \sigma_\mathrm{obs} ^2 - \frac{\Delta v_\mathrm{res}^2} {8 \,\mathrm{ln} 2},
\label{equ:velocity_disp}
\end{equation}
where $\Delta v_\mathrm{res}\sim0.16$\,\kms\ is the velocity resolution of the observations assuming a Gaussian response for the individual channels in the ALMA receiver (see \citealp{Koch2018} for effects of differing spectral response functions on the measured velocity dispersion).
Given that the line is resolved by a large number of channels ($\sigma_\mathrm{obs}^2/\sigma_\mathrm{res}^2\sim25$), this correction has a minor effect.  
Using $\sigma_\mathrm{obs,corr}$, we determined the contribution of the non-thermal motions to the velocity dispersion as
\begin{equation}
    \sigma_\mathrm{NT}^2 = \sigma_\mathrm{obs,corr}^{2} - \sigma_\mathrm{T}^{2} = \sigma_\mathrm{obs,corr}^{2} - \frac{k_\mathrm{B}T_\mathrm{kin}}{{m_\mathrm{obs}}},
\end{equation}
\noindent where $\sigma_\mathrm{NT}$ is the non-thermal velocity dispersion, $\sigma_\mathrm{T}$ is the thermal velocity dispersion of the observed molecule, and $m_\mathrm{obs}$ is the observed molecular mass (for \ntdp, $m_\mathrm{obs}=30\,m_\mathrm{H}$). The 
$T_\mathrm{kin}$ is the kinetic temperature of the gas, which we assume takes the same value as $T_\mathrm{dust}$ (10.4\,K; see Sect. \ref{sec_frag}), and $k_\mathrm{B}$ is the Boltzmann constant.

We examined this non-thermal contribution with respect to the sound speed, $c_\mathrm{s} = \sqrt{k_\mathrm{B}T_\mathrm{kin} / m_\mathrm{H} \mu_\mathrm{p}}$, where $\mu_\mathrm{p}$ is the mean molecular mass (2.37 for molecular gas at the typical interstellar abundance of H, He, and metals).
This is referred to as the one-dimensional sonic Mach number,\footnote{Note that here we have determined the sonic Mach number ($\mathcal M_\mathrm{s}\,=\,\sigma _\mathrm{NT} / c_\mathrm{s}$) using the one-dimensional velocity dispersion ($\sigma_\mathrm{NT}$). However, this value can be converted to the three-dimensional sonic Mach number by accounting for a factor of $3^{0.5}$: $\mathcal M_\mathrm{s,3D} = 3^{0.5} \mathcal M_\mathrm{s}$ (e.g. \citealp{palau15}).} or $\mathcal M_\mathrm{s}\,=\,\sigma _\mathrm{NT} / c_\mathrm{s}$. 
We find $\mathcal M_\mathrm{s}$ of 1.8 and 1.2 for C2c1a and C2c1b, highlighting that both cores appear to have either transonic turbulence or ordered global collapse motions.

We also calculated the total velocity dispersion, $\sigma_\mathrm{tot}$, which includes both thermal and non-thermal motions, following \citet{fuller_1992},
\begin{equation}
    \sigma_\mathrm{tot}^{2} = \sigma_\mathrm{obs,corr}^{2} + c_\mathrm{s}^2 - \sigma_\mathrm{T}^2 = \sigma_\mathrm{obs,corr}^{2} + k_\mathrm{B}T_\mathrm{kin} \left( \frac{1}{\bar{m}} - \frac{1}{m_\mathrm{obs}} \right),
\end{equation}
\noindent where $\bar{m}=m_\mathrm{H} \mu_\mathrm{p}$. 
We find $\sigma_\mathrm{tot}$ of 0.39\,\kms\ and 0.31\,\kms\ for C2c1a and C2c1b, which are used in the following stability analysis (Sects. \ref{sec_stab2} and \ref{sec_stab3}).

%
\section{Stability of C2c1}
\label{sec_stab}

%
\subsection{Thermal support}
\label{sec_stab1}

We determined the so-called Jeans mass, $M_\mathrm{J}$, which gives the maximum mass that can be supported by thermal pressure, and the Jeans length, $\lambda_\mathrm{J}$, which gives the minimum scale for fragmentation. The Jeans mass can be given as \citep{jeans_1902} 
\begin{equation}
M_\mathrm{J} = \frac{\pi^{5/2} c_\mathrm{s}^3}{6 G^{3/2} \rho^{1/2}}, 
\label{eq:jeansmass}
\end{equation}
where $\rho$ is the volume density of the core, and $G$ is the gravitational constant. 
We find $M_\mathrm{J}$ values of 0.05 and 0.13\,\msun, corresponding to $M/M_\mathrm{J}$ of $\sim$460 and 12 for C2c1a and C2c1b. 
These ratios of $M/M_\mathrm{J}>1$  show that the cores are unstable to gravitational collapse if not additionally supported. 
We estimated the corresponding Jeans length using 
\begin{equation}
\lambda_\mathrm{J} = c_\mathrm{s} \left (  \frac{\pi}{G \rho}  \right ) ^{1/2}.
\label{eq:jeanslength}
\end{equation}
We find $\lambda_\mathrm{J}$  values of 0.0035 and 0.0090\,pc for C2c1a and C2c1b (or 733 and 1900\,AU). 
Comparing these values to the projected radius of the cores, we find (2$r_\mathrm{eff})/\lambda_\mathrm{J}$ of around 8 and 3, highlighting that these Jeans-unstable cores could then fragment on size scales similar to the current observed core size scales. 

%
\subsection{Thermal and turbulent support}
\label{sec_stab2}

We then assessed the balance of the total kinetic energy, $E_{\rm kin}$, including both the thermal and turbulent pressure support, against the gravitational potential energy, $E_{\rm pot}$. 
These energy terms can be equated to produce the commonly used virial parameter $\alpha_{\rm vir}$ (e.g. \citealp{bertoldi_1992}). 
In the idealised case of a spherical core of uniform density supported by only kinetic energy (i.e. no magnetic fields), the virial parameter takes the form 
\begin{equation}
\alpha_{\rm vir} = a \frac{5 \sigma_\mathrm{tot}^{2} r_\mathrm{eff}}{G M},
\label{equ:virial}
\end{equation}
where $r_\mathrm{eff}$ is the effective radius of the core, $M$ is the (background-subtracted) mass of the core, and $\sigma_\mathrm{tot}$ is the line-of-sight velocity dispersion, that is, assuming the dispersion is a result of the thermal and turbulent broadening ($\alpha_{\rm vir}$ does not account for any systematic infall or outflow \LEt{ We reserve the use of slashes to denote ratios and instrument pairings and for use in equations. The use of "and/or" is acceptable. Kindly rephrase here and where marked.}motions; see e.g. \citealp{kauffmann_2013}). 
The factor $a$, which accounts both for systems with non-homogeneous and non-spherical density distributions and for a wide range of core shapes and density gradients, takes a value of $a = 2\,\pm\,1$ (see \citealp{bertoldi_1992}).
We find virial parameters of 0.1 and 0.7 for C2c1a and C2c1b, respectively, which would be indicative of them being bound and unstable to collapse (i.e. $\alpha_{\rm vir} < 2$). 

We also assessed the fragmentation of the cores using the total Jeans mass, $M_\mathrm{J,tot}$, which accounts for the contribution of both the thermal and non-thermal velocity dispersion (no infall or outflow \LEt{ slash.}motions). 
This can be calculated by substituting $\sigma_\mathrm{tot}$ for $c_\mathrm{s}$ in Eq.\,\ref{eq:jeansmass} (e.g. \citealp{palau15}).\LEt{ Verify that your intended meaning has not been changed.} 
We find $M_\mathrm{J,tot}$ of 0.4 and 0.5\,\msun, or $M/M_\mathrm{J,tot}$ of 55 and 3, for C2c1a and C2c1b. 
These total Jeans masses  are typically factors of a few higher than when only the thermal support ($M_\mathrm{J}$) is accounted for (see Sect. \ref{sec_stab1}).
\LEt{ Verify that your intended meaning has not been changed.}However, as shown by the virial parameter, values of $M/M_\mathrm{J,tot}>1$ highlight that the cores are still likely to collapse and/or fragment unless even further supported.

%
\subsection{Thermal, turbulent, and magnetic support}
\label{sec_stab3}

We assessed the relative importance of the magnetic field in preventing gravitational collapse. 
To do so, we calculated the virial parameter that includes the magnetic field contribution \citep{pillai_2011}, 
\begin{equation}
\alpha_{\rm B, vir} = a \frac{5 r_\mathrm{eff}}{G M} \left ( \sigma_\mathrm{tot}^{2} - \frac{ v_\mathrm{A}^{2} }{6} \right ),
\label{equ:virialmag}
\end{equation}
where the Alfv\'en velocity is $v_\mathrm{A} = B (\mu_0 \rho)^{-1/2}$, in which $B$ is the magnetic field strength and $\mu_0$ is the permeability of free space (again, no infall or outflow \LEt{ slash.}motions are considered in the velocity dispersion).\footnote{This expression is valid for SI units; in cgs, the permeability of free space is unity and, hence, removed from the expression.\LEt{ Verify that your intended meaning has not been changed.}} 
Here then we ask how much magnetic field pressure is required in addition to turbulence and thermal pressure to support the cores against gravity. 
To answer this question, we set $a=2$ and solved Eq.\,\ref{equ:virialmag} for $B$ for $\alpha_{\rm B,vir}<2$ (Sect. \ref{sec_stab2}). 
We find that the magnetic field strengths required for stability are 10.0\,mG and 1.0\,mG for C2c1a and C2c1b.

We compared these estimates to the \citet{crutcher_2010} relation linking the magnetic field strength (determined from Zeeman splitting) and volume density (also see \citealp{Liu2022}), 
\begin{equation}
B_\mathrm{med} \approx \frac{1}{2} B_0 \left( \frac{n(\mathrm{H})}{n_0} \right) ^{2/3} \approx \frac{1}{2}  B_0 \left( \frac{2n_\mathrm{H_2}}{n_0} \right) ^{2/3},
\label{equ:Bmed}
\end{equation}
where $B_\mathrm{med} = B_\mathrm{max}/2$, and for $n(\mathrm{H})$\,$>$\,$n_0$, where $n_0 = 300$\,\cmcb, $n(\mathrm{H})=2n_\mathrm{H_2}$, and $B_0 = 10$\,$\mu$G. 
We find $B_\mathrm{med}$ values of 14.4\,mG and 4.3\,mG, or $B/B_\mathrm{med}$ of 0.7 and 0.2, for C2c1a and C2c1b. 
This shows that the magnetic field required for the additional support against gravitational collapse could then be more than achieved if these cores follow the \citet{crutcher_2010} relation, which is broadly consistent with the typical magnetic field strengths observed within molecular clouds \citep{pillai_2015, pillai_2016, soam_2019, tang_2019}. 

Lastly, we compared the measured core properties to predictions of the turbulent core model \citep{mckee_2003}. 
The mass-weighted average velocity dispersion of a virialised core, including pressure equilibrium with its surroundings, is given in the fiducial case by \citep{mckee_2003, Tan2013}\footnote{The fiducial case presented here is outlined in Eq. 18 of \citet{mckee_2003}, where the fiducial values are taken from their Eq. 12 (also see \citealp{Kong2018}; their Eq. 2).}
\begin{equation}
\sigma_\mathrm{c,vir} = 1.09 \left( \frac{\phi_\mathrm{B}}{2.8} \right)^{-3/8} \left( \frac{M}{60 \,\mathrm{M_\odot}} \right)^{1/4} \left( \frac{\Sigma_\mathrm{cl}}{1 \,\mathrm{g\,cm^{-2}}} \right)^{1/4} \mathrm{km\,s^{-1}}
\label{equ:sigma_c,vir}
,\end{equation}
where $\phi_\mathrm{B} = 1.3 + 1.5 \mathcal{M}_\mathrm{A}^{-2} = 2.8$ is a dimensionless parameter that accounts for the effects of magnetic fields ($\mathcal{M}_\mathrm{A}=3^{0.5}\sigma_\mathrm{tot}/v_\mathrm{A}=1$ is the fiducial Alfv\'en Mach number), and $\Sigma_\mathrm{cl} = 0.6\mathrm{\,g\,cm^{-2}}$ is the mass surface density of the large-scale C2c1 core taken from the ALMA Infrared-Dark Cloud (IRDC) survey (see \citealp{Barnes2021_almairdc}).
We find $\sigma_\mathrm{c,vir}$ of 0.70\,\kms\ and 0.35\,\kms\ for C2c1a and C2c1b, or $R_\sigma=\sigma_\mathrm{tot}/\sigma_\mathrm{c,vir}$ of 0.56 and 0.87.
These values of $R_\sigma=\sigma_\mathrm{tot}/\sigma_\mathrm{c,vir}<1$ obtained with the above fiducial values would suggest that the core is in a sub-virial state, and on the verge of undergoing collapse because of the lack of sufficient internal pressure support.
However, if magnetic fields play a more important role, then this is represented by a smaller Alfv\'en Mach number (i.e. sub-Alfv\'en turbulence) and a larger value of $\phi_\mathrm{B}$.
For example, using the previously estimated values of $B_\mathrm{med}$ gives $\sigma_\mathrm{c,vir}(B_\mathrm{med})$ of 0.26\,\kms\ and 0.14\,\kms, and $\sigma_\mathrm{tot}/\sigma_\mathrm{c,vir}(B_\mathrm{med})$ of 1.5 and 2.3, which would imply that the cores are close to virial equilibrium and are contracting relatively slowly compared to free-fall collapse.

%
\section{Pre-stellar versus protostellar}
\label{sec_sf}

\begin{figure}
    \centering
        \includegraphics[width=\columnwidth]{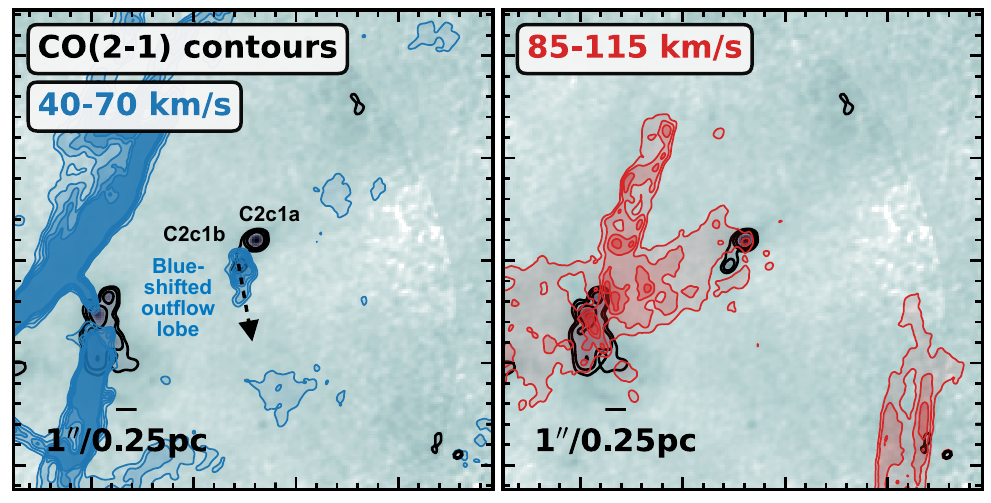}
    \caption{{Evidence for a molecular outflow towards the core.} We show in contours the CO\,(2-1) emission integrated in the ranges 40-70\,\kms\ (blue contours) and 85-115\,\kms\ (red contours). The background grey-scale image shown in both panels is the 1\,mm continuum emission (see Fig.\,\ref{fig1}). Highlighted is the associated blueshifted outflow lobe (no redshifted lobe is obvious due to line-of-sight contamination at higher velocities). Shown in the bottom right is the beam size of the observations. Figure\,\ref{fig_outflow_co} shows channel maps of the CO emission across the region.}
    \label{fig_cooutflow}
\end{figure}

The next step is then to determine whether the cores are truly quiescent or show signs of active star formation, which is a key factor in distinguishing between the theories of massive star formation (Sect. \ref{sec_int}).

Infrared emission could highlight any embedded protostars within IRDCs, due to the heated dust emission, polycyclic aromatic hydrocarbon emission, or molecular line emission (e.g. \citealp{Chambers2009}). 
We checked {\it Spitzer} (e.g. 3.6 to 8\micron, and 24\micron; \citealp{churchwell_2009, carey_2009}) and {\it Herschel} (70\micron, \citealp{molinari_2010}) data, and confirmed that there are no infrared point sources within any of the bands towards the position of the C2c1 core. 
The completeness of these surveys at the cloud distance is approximately a few solar masses\LEt{ Units should be spelled out in full when not following a numeral.\ Please verify this is the correct definition.}, depending on the evolutionary stage of the young stellar object (see efforts to classify young stellar objects in similar IRDCs, e.g. \citealp{nguyen_2011}). 
The lack of infrared emission is consistent with them being cold, dense, and quiescent.

In addition, we checked centimetre-radio observations from the THOR \LEt{ Consider defining.}project \citep{Beuther2016, Wang2020} and find no significant continuum or radio recombination line emission in the C2 clump region, which would be indicative of a lack of embedded \HII\ regions (i.e. from free-free emission).
The integrated continuum flux density within a beam-sized aperture  (25\arcsec) towards the core C2c1 at 1.42\,GHz is $\sim$\,10\,mJy (calculated using the THOR-VLA data combined with the Green Bank Telescope (GBT)\LEt{ Consider defining.}.
More recently, Wang et al. (in prep) observed this region with higher-resolution and higher-sensitivity MeerKAT \LEt{ Consider defining.}L-band observations. The 1.4 GHz continuum image has a resolution of about 6.5 arcsec and rms 65\,$\mu$Jy/beam, and no compact radio sources are seen in the clump. For a box encompassing the starless core, the maximum detected flux is 6\,mJy, in broad agreement with the VLA observations.
Using the conversion from \citet{mezger_1967}, these values give a Lyman-continuum photon rate of $\sim$\,10$^{46}$\,s$^{-1}$, which, based on the stellar models from \citet{Smith2002}, gives a zero-age main sequence spectral type of at most B1.5.
Hence, we can rule out the presence of any O-type or early B-type stars towards this region.

Moreover, assuming thermal equilibrium between gas and dust, \citet{Wang2012} used the measured temperature of the C2c1 core ($\sim$\,10\,K) to estimate that an at most $\sim$\,1\,$\mathrm{L_\odot}$ protostar could be present within the core.  
In addition, \citet{Wang2012} observed H$_2$O and CH$_3$OH masers with VLA and found a few in the clump, but not near the C2c1 core.
All this together again rules out the presence of any significant evolved star formation.

Several works have, however, noted the presence of a weak molecular outflow towards the region (e.g. \citealp{Wang2011,Kong2019}). 
We investigated this using new higher-sensitivity CO\,(2$-$1) observations at an angular resolution of $\sim$\,0.3\arcsec (see Fig.\,\ref{fig_cooutflow}) -- almost an order of magnitude higher resolution than previous studies.
We find a weak outflow signature between 40\,\kms\ and 70\,\kms, which appears to be reminiscent of a single outflow lobe associated with the core that is blueshifted with respect to the systemic velocity (as seen in \ntdp; $\sim$\,80\,\kms).
\citet{Wang2011} also determined that C2c1 is driving a weak outflow with a redshifted lobe between 86 and 91\,\kms, as seen in CO (3$-$2) emission. 
The identification of such a component is complicated in the higher-resolution CO (2--1) data presented in this work due to the presence of unassociated emission along the line of sight at these higher velocities and imaging artefacts (between 85 and 95\,\kms; see Fig.\,\ref{fig_outflow_co}). 
\citet{Wang2011} estimate that this core \LEt{ Verify that your intended meaning has not been changed.}has the smallest mass outflow rate in the region (0.73$\times$\,10\,$^{-5}$\,\msun\,yr$^{-1}$).\ It is also worth highlighting that no associated SiO outflow has been identified towards this region \citep{LiuM2020}.
Therefore, the outflow is then comparatively weak, which could be indicative of its relative youth.
Interestingly, our higher-resolution images show that the outflow begins at the position of the C2c1b core, and that the orientation does not clearly connect to the C2c1a core (see Appendix \ref{sec_app_outflows} for more detail channel maps of the CO emission). 
These new data \LEt{ these what? images? facts?}could suggest that this outflow is not related to the C2c1a core (as previously suggested), but rather to the C2c1b core.
This is an important distinction because in this case, despite its significant mass, C2c1a would remain completely devoid of any active star formation.\LEt{ Verify that your intended meaning has not been changed.}

\section{Chemistry}

\begin{figure*}
    \centering
        \includegraphics[width=\textwidth]{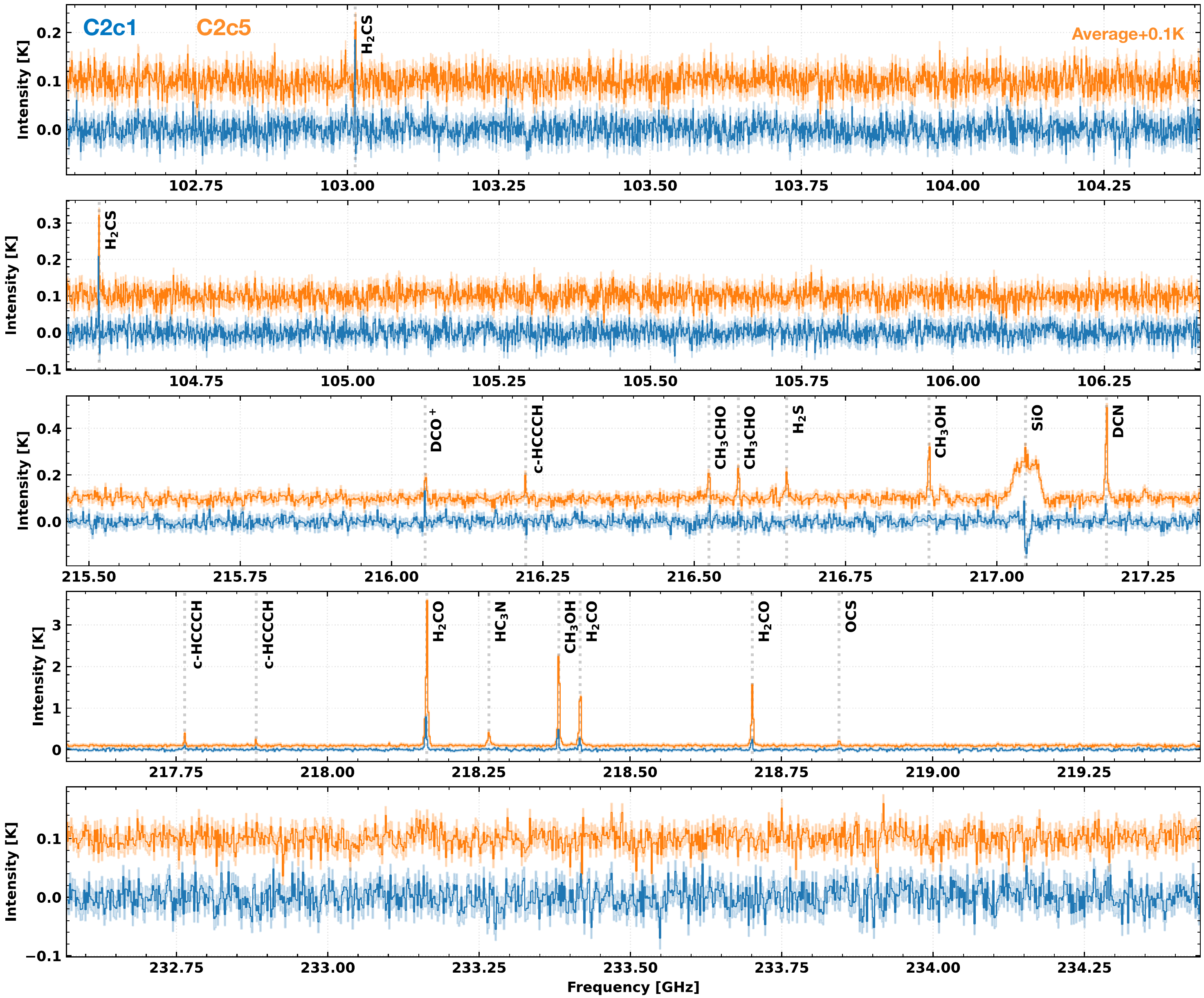}
    \caption{{Spectrum taken across the 3\,mm \citep{Barnes2021_almairdc} and 1\,mm spectral windows towards the C2c1 (blue) and C2c5 (orange) cores.} Shown as coloured lines is the spectrum averaged within the boundary of the cores, and the coloured shaded regions indicate the uncertainty (rms). The spectrum of the C2c5 core has been offset by +0.1K to allow comparison. Several prominent lines within the spectral windows, which are summarised in Table\,\ref{tab2}, are labelled.}
    \label{fig_chemspec}
\end{figure*}

 Chemistry can also be used as an informative proxy for the evolutionary state of a star-forming region.
For example, the number of detected lines and their relative strengths can be used to differentiate the pre-stellar or protostellar nature of dense cores (e.g. \citealp{Nony2018}).     
To investigate this we show the average broadband spectrum across the C2c1 region in Fig.\,\ref{fig_chemspec} (blue).
In Fig.\,\ref{fig_chemspec}, the 3\,mm (102.55 to 104.35\,GHz and 104.55 to 104.35\,GHz ) data are taken from the ALMA-IRDC survey \citep{Barnes2021_almairdc}, and the 1\,mm data (215.45 to 217.35\,GHz, 217.55 to 219.35\,GHz, and 232.55 to 234.35\,GHz) are those presented in this work (see Sect. \ref{sec_obs}).  
We see that the C2c1 region contains only a limited number of weak lines, which we identified using the {\sc casaviewer} tool (making use of the Splatalogue database; \citealp{Remijan2007}).
The detected molecules in emission include DCO$^+$ and DNC, which are indicative of the core's early evolutionary stage, as well as H$_2$CS, H$_2$CO, and CH$_3$OH with $T_\mathrm{up}=20-50$\,K, which indicates relatively cold temperatures (see Table\,\ref{tab2}).  
In Fig.\,\ref{fig_chemspec}, we compare these results to a more actively star-forming core within the region (C2C5; see \citealp{Barnes2021_almairdc}). 
We see that lines within C2c1 are both systematically weaker and fewer in number compared to C2C5 (also see \citealp{Zhang2015}).
Indeed, C2C5 has a spectrum more comparable to that of hot core candidates (see \citealp{Herbst2009}).
Of particular note is the strong absorption feature seen in SiO\,(5-4) towards C2c1 that is clearly seen in emission towards C2C5. 
This is a strong differentiator for the star formation activity of the two cores and further indicates that C2c1 is young, a conclusion that is also in agreement with the lack of SiO outflows previously found in the region.

\begin{table}
\caption{{Summary of lines detected towards the cores (see Fig. \,\ref{fig_chemspec}).} We tabulate the line species, transition, rest frequency, and upper energy level, which are taken from the Splatalogue database \citep{Remijan2007}.}              
\label{tab2}      
\centering                                      
\begin{tabular}{llcc}
\hline \hline
Line & Transition & $\nu$ [$\mathrm{MHz}$] & $T_\mathrm{up}$ [K] \\
 \hline
H$_2$CS & 3(0,3)-2(0,2) & 103040 & 9.9 \\
H$_2$CS & 3(1,2)-2(1,1) & 104617 & 23.2 \\
DCO$^+$ & J=3-2 v=0 & 216112 & 20.7 \\
c-HCCCH & 3(3,0)-2(2,1) $v$=0 & 216278 & 19.5 \\
CH$_3$CHO & 11(1,10)-10(1,9) E $v_t$=0 & 216581 & 64.9 \\
CH$_3$CHO & 11(1,10)-10(1,9) A -- $v_t$=0 & 216630 & 64.8 \\
H$_2$S & 2(2,0)-2(1,1) & 216710 & 84.0 \\
CH$_3$OH & 5(1,4)-4(2,2) & 216945 & 55.9 \\
SiO & 5-4 $v_t$=0 & 217104 & 31.3 \\
DCN & J=3-2 $v$=0 & 217238 & 20.9 \\
c-HCCCH & 6(1,6)-5(0,5) $v$=0 & 217822 & 38.6 \\
c-HCCCH & 5(1,4)-4(2,3) $v$=0 & 217940 & 35.4 \\
H$_2$CO & 3(2,1)-2(0,2) & 218222 & 21.0 \\
HC$_3$N & J=24-23 $v$=0 & 218324 & 131.0 \\
CH$_3$OH & 4(2,2)-3(1,2) $v_t$=0 & 218440 & 45.5 \\
H$_2$CO & 3(2,2)-2(2,1) & 218475 & 68.1 \\
H$_2$CO & 3(2,1)-2(2,0) & 218760 & 68.1 \\
OCS & 18-17 $v$=0 & 218903 & 99.8 \\
\hline
\end{tabular}
\end{table}

\section{Discussion and summary}

Our analysis shows that the C2 core is a very interesting region in the context of massive star formation. 
We find that at the end of the chain of roughly equally spaced actively star-forming cores ($\sim$\,0.15 pc; \citealp{Zhang2009, Wang2011}) sits another massive, yet relatively quiescent, core -- C2c1. 
This \LEt{ Consider specifying what you mean by "this" - This placement? Its high mass?}indicates a potential role for filamentary clouds (see e.g. \citealp{Jimnez-Serra2014, Henshaw2013, Henshaw2014, Barnes2018}) in regulating the birth of massive pre-stellar cores.

High-angular-resolution (0.2\arcsec\ or 0.05\,pc) dust continuum observations show that this region fragments into two cores, C2c1a and C2c1b, which have considerable masses of 23\,\msun\ and 2\,\msun, respectively. 
Our stability analysis shows that both cores are highly unstable to gravitational collapse ($\alpha_{\rm vir}$ of 0.1 and 0.7) unless supported by magnetic fields of around 1 to 10\,mG in strength.
These values are high but broadly consistent with the typical magnetic field strengths observed within dense molecular clouds \citep{pillai_2015, pillai_2016, soam_2019, tang_2019}. 
Lastly, we find that there is a weak CO outflow that appears to be associated with C2c1b only. The more massive core, C2c1a, remains completely devoid of any star formation signatures and does not appear to be coincident with the chemistry observed towards more evolved hot cores.    
Overall, we find that both cores are good targets for studying the early stages of massive star formation.
They also offer an interesting comparison to other massive core candidates (e.g. \citealp{Bontemps2010, Duarte-Cabral2013, Wang2014, Cyganowski2014, Sanhueza2017, Nony2018, Louvet2019}), particularly when focusing on core candidates identified within Cloud C that have been analysed with methods similar to those used in this work \citep{Tan2013, Tan2016, Kong2018}.
\LEt{ Verify that your intended meaning has not been changed.}

\citet{Tan2013} used early (cycle 0) ALMA observations of $\rm N_2D^+$(3-2) towards IRDCs to identify the C1-S and C1-N cores as candidate massive pre-stellar cores, with masses of $16^{+34}_{-7}$\,\msun\ and $63^{+130}_{-27}$\,\msun, respectively, based on their 1.3\,mm dust continuum emission. 
Following up from this, \citet{Tan2016} and \citet{Kong2018} presented higher-resolution observations of the C1-S region, identifying two early stage protostellar cores in the vicinity; however, the main C1-S starless core was resolved as a distinct spatial and kinematic feature in its $\rm N_2D^+$ emission and with a millimetre-continuum derived mass of $59^{+123}_{-27}$\,\msun\ within a radius of 0.045\,pc. \LEt{ Verify that your intended meaning has not been changed.}
Their dynamical analysis of the cores based on the velocity dispersion measured via $\rm N_2D^+$(3-2) found moderately sub-virial conditions, and $R_\sigma=0.61^{+0.88}_{-0.44}$ for C1-N and $0.34^{+0.46}_{-0.27}$ for C1-S; the corresponding virial parameters are $\alpha_\mathrm{vir} = 0.98^{+2.4}_{-0.42}$ and $0.068^{+0.15}_{-0.03}$, respectively. 
These values are similar to those calculated for C2c1a and C2c1b (0.56 and 0.87, respectively), also suggesting that they are on the verge of collapse (Sect. \ref{sec_stab3}).

In addition, \citet{Kong2017} surveyed 32 IRDC regions and identified $\rm N_2D^+$(3-2) cores with the same method. 
The most massive core in this sample, C9A, has a mass of $M_{\rm c,mm}=69.7^{+146}_{-31.7}$\,\msun\ and $R_\sigma=0.71^{+0.89}_{-0.59}$, while the next most massive is B1A, with $M_{\rm c,mm}=4.9^{+10.3}_{-2.2}$\,\msun\ and $R_\sigma=0.85$, again similar to C2c1a and C2c1b. 
Despite these being particularly promising candidates for massive pre-stellar cores, their analysis is complicated by their proximity, at least in projection, to protostellar sources \citep{Kong2018}.

In combination with the above studies, our results indicate that Cloud C is a particularly rich candidate for studying the early stages of massive star formation. 
In particular, we find that one core, C2c1a, remains massive enough to form a high-mass star at a radius of around 0.01\,pc (or $\sim$\,2000\,AU) and yet still appears to be in a pre-stellar evolutionary phase. 
Such an object could be reminiscent of a massive starless (monolithic) core.\ The definitive identification of such cores has proven elusive, making C2c1a an exciting prospect for further studies.\LEt{ Verify that your intended meaning has not been changed.} 



\begin{acknowledgements}

We are grateful to the anonymous referee for their constructive and detailed suggestions, which helped significantly improve the quality of this paper.
We would like Henrik Beuther for their insightful discussions and comments on the draft. 
ATB and FB would like to acknowledge funding from the European Research Council (ERC) under the European Union’s Horizon 2020 research and innovation programme (grant agreement No.726384/Empire).
J.D.H gratefully acknowledges financial support from the Royal Society (University Research Fellowship)
JCT acknowledges support from ERC Advanced Grant MSTAR and NSF grant AST-2009674.
KW acknowledges support from the National Science Foundation of China (11973013), China Manned Space Project (CMS-CSST-2021-A09), National Key R\&D Program of China (2022YFA1603102).

\end{acknowledgements}

\bibliographystyle{aa}
\bibliography{references}

\begin{appendix}

\section{Integrated intensity maps of N$_2$D$^+$}
\label{sec_app_n2d}

\begin{figure*}
    \centering
        \includegraphics[width=0.95\textwidth]{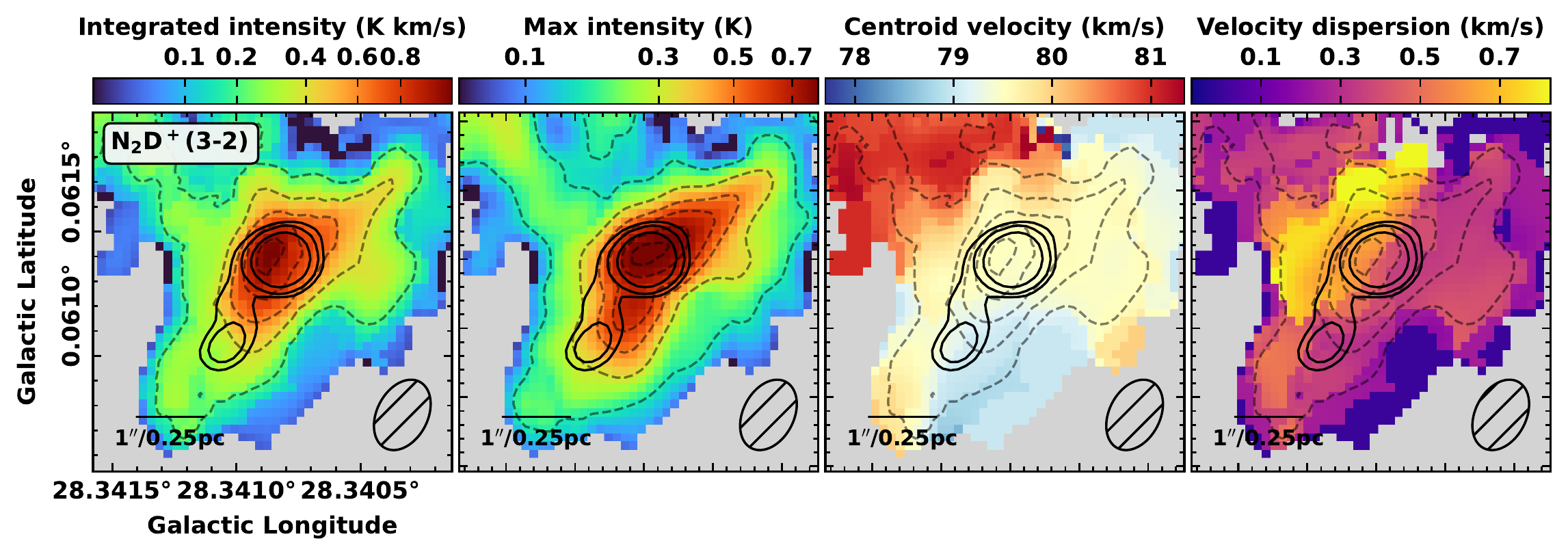}
    \caption{{Distribution of \ntdp\,(3-2) emission across the core C2c1 region.} {\it (panels from left to right)} Integrated intensity, peak intensity, intensity-weighted centroid velocity, and intensity-weighted velocity dispersion. Overlaid as solid black contours is the ALMA 1\,mm dust continuum at $0.3$\arcsec\ resolution (see Fig.\,\ref{fig1}). The dashed contours in the first, third, and fourth panels are the integrated intensity, and in the second panel they are the maximum intensity. Shown on the right of the panels is the beam size, and a scale bar is shown in the lower left of all panels.}
    \label{fig_nthp_moms}
\end{figure*}

\begin{figure*}
    \centering
        \includegraphics[width=\textwidth]{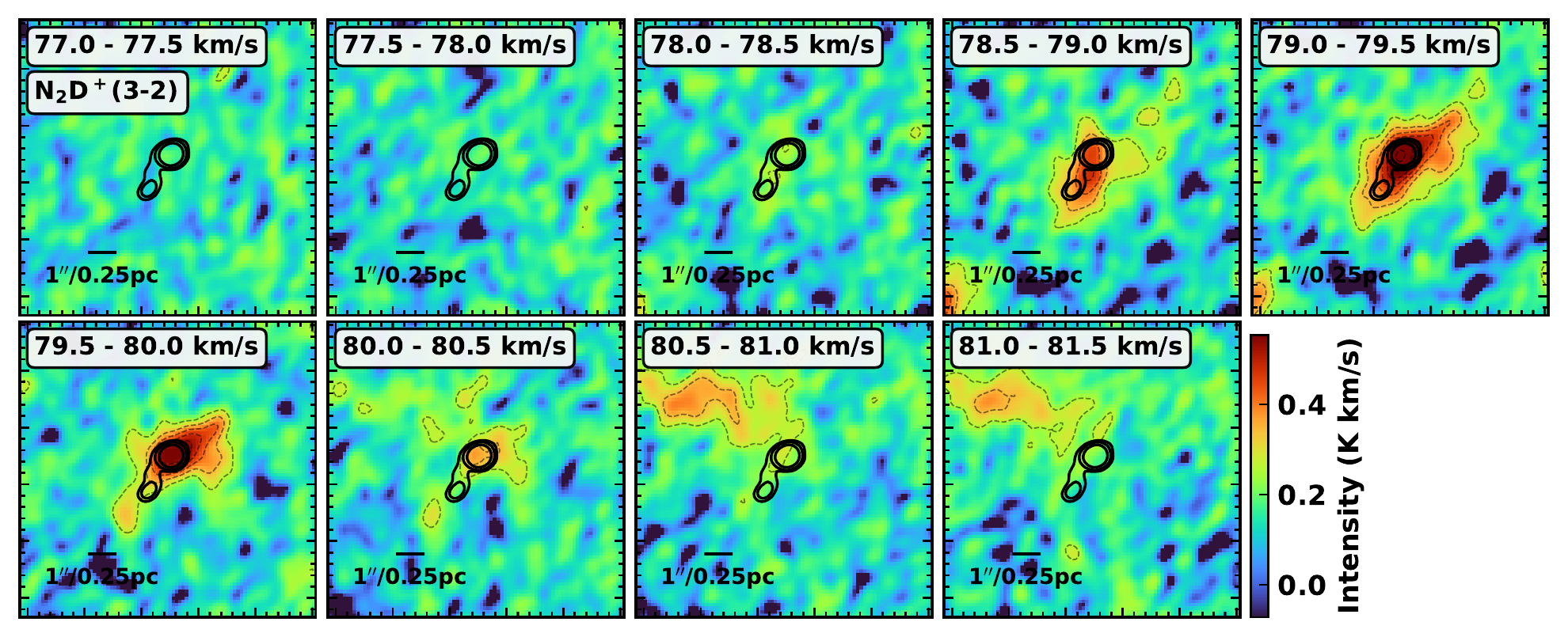}
    \caption{{Channel maps of the N$_2$D$^+$\,(3--2) emission across the core region.}  Overlaid as contours is the ALMA 1\,mm dust continuum at $0.3$\arcsec\ resolution (see Fig.\,\ref{fig1}). A scale bar is shown in the lower left of all panels.}
    \label{fig_nthp_channelmaps}
\end{figure*}


In Fig.\,\ref{fig_nthp_moms}, we show moment maps (integrated intensity, max intensity, intensity-weighted centroid velocity, and intensity-weighted velocity dispersion) of the \ntdp\,(3-2) emission. 
We see that the \ntdp\ emission is widespread across both cores as seen in the dust continuum, even bridging the region between the cores. 
This indicates that the gas surrounding the cores is also cold and could contribute to accretion.
We also see that there is a velocity gradient across the core region, from low velocities at C2c1b to high velocities at C2c1a, which could indicate that the system is rotating.
This velocity gradient can also be seen in the channel maps presented in Fig.\,\ref{fig_nthp_channelmaps}, which shows the \ntdp\ emission integrated in velocity bins of 0.5\,\kms.
Interestingly, we see that there is an emission feature that extends to the north of the system in the velocity bins of 80.5 -- 81.5\,\kms, which can also be seen as a sharp cut in the velocity field map (Fig.\,\ref{fig_nthp_moms}).
If this emission is connected to the core, this could suggest that dense molecular gas is still being accreted onto the core envelope.  
An in-depth analysis of inflow and the accretion signature will be performed in a future work.


\section{Outflows}
\label{sec_app_outflows}

\begin{figure*}
    \centering
        \includegraphics[width=\textwidth]{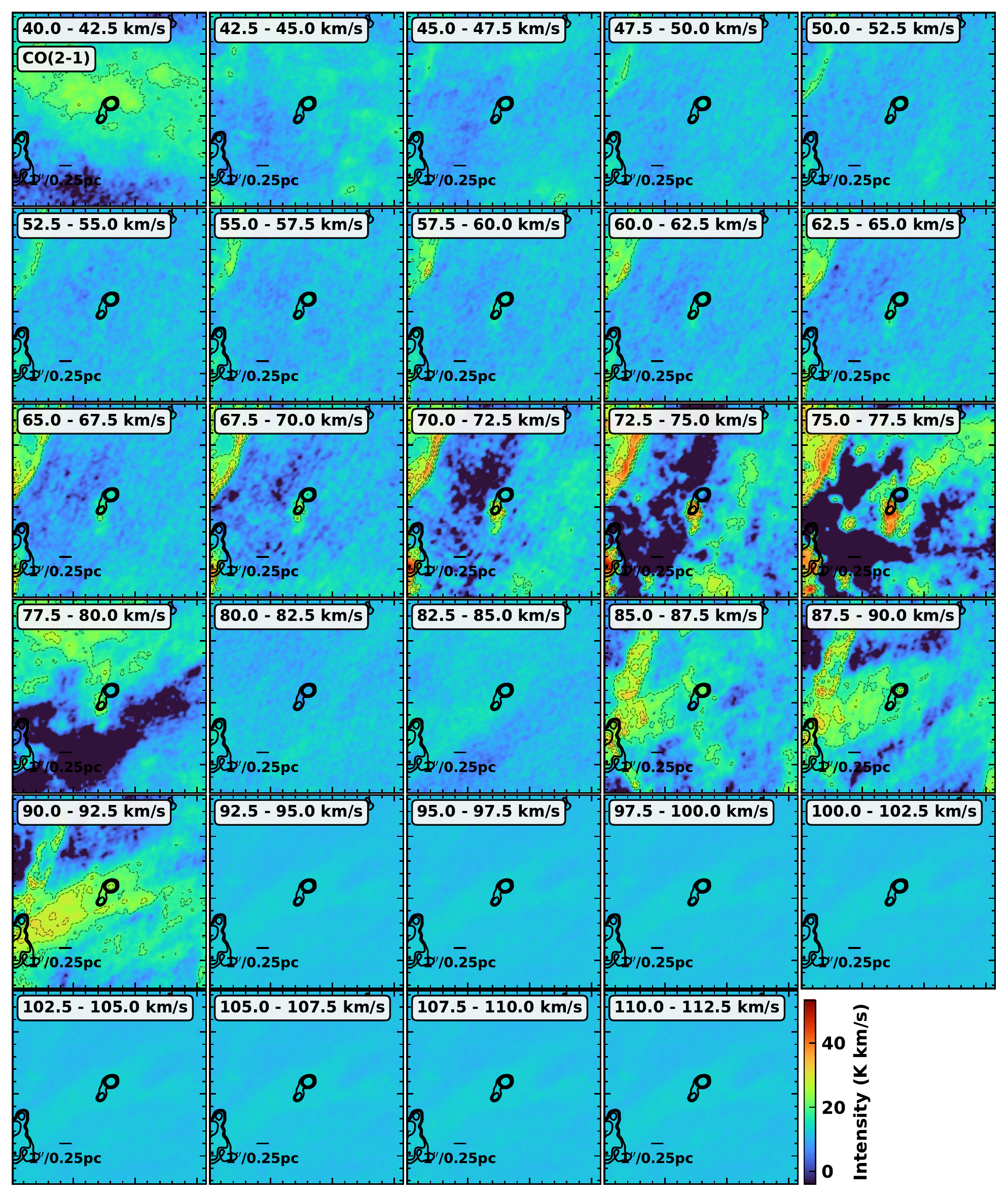}
    \caption{{Channel maps of the CO\,(2--1) emission across the core region.}  Overlaid as contours are the ALMA 1\,mm dust continuum at $0.3$\arcsec\ resolution (see Fig.\,\ref{fig1}). A scale bar is shown in the lower left of all panels.}
    \label{fig_outflow_co}
\end{figure*}

In Fig.\,\ref{fig_outflow_co} we show the CO\,(2-1) emission integrated in velocity bins of 2.5\,\kms\ from 40\,\kms\ to 112.5\,\kms. 
The outflow feature can be seen towards core C2c1b between velocities of 55\,\kms\ and 77.5\,\kms. 
We do not find a similar feature at higher velocities that extends across such a large velocity range ($\sim$\,20\,\kms).
Indeed, the identification of a redshifted counterpart to this outflow (e.g. from around 70\,\kms\ to 90\,\kms) is complicated by the presence of imaging artefacts.
That said, there is some emission in the direction where the redshifted lobe would be expected within the 75\,\kms\ to 77.5\,\kms\ channel. 
Yet, it is not clear if it is associated with the extended emission along the line of sight that can also be seen at higher velocities (85\,\kms\ to 92.5\,\kms).

\end{appendix}

\end{document}